\documentstyle[epsfig,prl,twocolumn,aps]{revtex} 
\begin{document}
\def\beq{\begin{equation}}
\def\eeq{\end{equation}}
\def\fig#1{{Fig. (\ref{#1})}}
\def\bea{\begin{eqnarray}}
\def\eea{\end{eqnarray}}
%%\rightline{hep-ph/9712487}
\def\thefootnote{\fnsymbol{footnote}}
\newcommand{\ptmis}{{ {\rm p} \hspace{-0.53 em} \raisebox{-0.27 ex} {/}_T }}
 \def\mathrm#1{\mbox{\rm #1}}
\def\bold#1{\setbox0=\hbox{$#1$}%
     \kern-.025em\copy0\kern-\wd0
     \kern.05em\copy0\kern-\wd0
     \kern-.025em\raise.0433em\box0 }
\def\21{$SU(2) \otimes U(1)$}
\newcommand{\matriz}{\left[\begin{array}} 
\newcommand{\finmatriz}{\end{array}\right]} 
\def\frad#1#2{\frac{\displaystyle{#1}}{\displaystyle{#2}}}
\def\frap#1#2{{\hbox{$\frac{#1}{#2}$}}}
\def\half{{\textstyle{1 \over 2}}}
\def\etal{\hbox{\it et al., }}
\def\eq#1{{eq. (\ref{#1})}}
%%%%%%%%%%%%%%%%%%%%%%%%%%%%%%%%%%
\def\sm{\hbox{Standard Model }}
\def\VEV#1{\left\langle #1\right\rangle}
\def\lsim{\raise0.3ex\hbox{$\;<$\kern-0.75em\raise-1.1ex\hbox{$\sim\;$}}}
\def\gsim{\raise0.3ex\hbox{$\;>$\kern-0.75em\raise-1.1ex\hbox{$\sim\;$}}}
\def\mpl#1#2#3{          {\it Mod. Phys. Lett. }{\bf #1} (19#2) #3}
\def\np#1#2#3{           {\it Nucl. Phys. }{\bf #1} (19#2) #3}
\def\nps#1#2#3{          {\it Nucl. Phys. B (Proc. Suppl.) }
    {\bf #1} (19#2) #3}
\def\pl#1#2#3{           {\it Phys. Lett. }{\bf #1} (19#2) #3}
\def\ppnp#1#2#3{           {\it Prog. Part. Nucl. Phys. }{\bf #1} (19#2) #3}
\def\pr#1#2#3{           {\it Phys. Rev. }{\bf #1} (19#2) #3}
\def\prep#1#2#3{         {\it Phys. Rep. }{\bf #1} (19#2) #3}
\def\prl#1#2#3{          {\it Phys. Rev. Lett. }{\bf #1} (19#2) #3}
\def\zp#1#2#3{          {\it Z. Phys. }{\bf #1} (19#2) #3}
\twocolumn[\hsize\textwidth\columnwidth\hsize\csname @twocolumnfalse\endcsname
\title{ Neutrino Mass and Missing Momentum Higgs Boson Signals }
%%in $e^+ e^-$ Collisions}
%
\author{ M. A. D\'\i az,
M. A. Garc{\'\i}a-Jare\~no, D. A. Restrepo and
 J. W. F. Valle }
\address{Instituto de F\'{\i}sica Corpuscular - IFIC/CSIC,
             Departament de F\'{\i}sica Te\`orica \\
             Universitat de Val\`encia, 46100 Burjassot, 
             Val\`encia, Spain\\
	     http://neutrinos.uv.es}
\date{\today} 
\maketitle
\begin{abstract} 
In the simplest scheme for neutrino masses invoking a triplet of Higgs
scalars there are two CP-even neutral Higgs bosons $H_i$ (i=1,2) and
one massive pseudoscalar $A$.  For some choices of parameters, the
lightest $H_1$ may be lighter than the Standard Model Higgs boson. If
the smallness of neutrino mass is due to the small value of the
triplet expectation value, as expected in a seesaw scheme, the Higgs
bosons may decay dominantly to the invisible neutrino channel. We
derive limits on Higgs masses and couplings that follow from LEP I
precision measurements of the invisible Z width.  
\end{abstract}
\pacs{14.80.Cp, 13.85.Qk}
\vskip2pc]

Neutrino mass constitutes one of the deepest open issues in the
Standard Model (SM) of particle physics, which now finds some
observational support \cite{revnu}. Neutrino masses
in the few eV range may also be crucial in explaining the
large scale structure of the universe. In many \21 extensions of the SM
neutrino masses are generated from the spontaneous violation of a
global lepton number symmetry leading to the existence of a physical
Goldstone boson - the majoron \cite{CMP}. In such models the majoron
acts as the tracer of the neutrino mass generation mechanism and may
have, depending on the details of the model, many interesting
phenomenological implications \cite{fae}.  If the breaking of lepton
number occurs at the weak scale the lightest Higgs boson can
decay dominantly into the weakly interacting majorons
\cite{HJJ}. Since these escape detection, this decay is called
invisible and has as signature missing momentum in the reaction.  Here
we consider a more direct way in which neutrino mass physics can show
up as a missing momentum Higgs boson signature.  As our illustrative
model we consider the simplest triplet model for generating neutrino
masses as first proposed, in the majoron-less form \cite{2227}. The
model contains a complex $SU(2)$ triplet of scalar bosons $\Delta$, in
addition to the standard Higgs doublet $\phi$
\begin{equation}
\phi=\matriz{c}\phi^0\\
\phi^-\finmatriz, \:
\Delta=\matriz{cc} \Delta^0&\Delta^+/\sqrt2\\
\Delta^+/\sqrt2&\Delta^{++}\finmatriz,
\label{higgs}
\end{equation}
where we have used the $2\times 2$ matrix notation for the Higgs
triplet.  Apart from obvious terms in $\phi$ and $\Delta$, the scalar
potential contains two mixed quartic terms and one tri-linear term
given as \def\tr{{\rm tr}\,} 
\beq
\lambda_3\phi^\dagger\phi\,\tr(\Delta^\dagger\Delta)
+\lambda_5(\phi^\dagger\Delta^\dagger\Delta\phi)
-\frac{\kappa}{\sqrt2}(\phi^T\Delta\phi+{\rm h.c.})
\label{eq:GenPot}
\eeq
where the $\lambda$'s are dimensional-less couplings and $\kappa$ has
mass dimension and should lie at the weak scale or less.  Note that
the smallness $\kappa$ is natural according to 't Hooft's criterion,
since the symmetry of the model increases as $\kappa \to 0$. We assume
that the fields $\phi$ and $\Delta$ acquire nonzero real vacuum
expectation values (VEVS) $v_2$ and $v_3$, respectively.  It is easy
to verify explicitly that, for many choices of its parameters, the
potential has indeed minima for nonzero values of $v_2$ and $ v_3$.
According to this, we shift the fields in the following way
\begin{eqnarray}
\phi^0&=&\frad{v_2}{\sqrt{2}}+\frad{R_2+iI_2}{\sqrt{2}}
\nonumber\\
\Delta^0&=&\frad{v_3}{\sqrt2}+\frad{R_3+iI_3}{\sqrt{2}}
\label{shift}
\end{eqnarray}
Notice that the existence of a cubic term in the scalar potential breaks
explicitly the lepton number symmetry, avoiding the
Gelmini--Roncadelli triplet majoron \cite{GelRon,GGN} - now ruled out
experimentally by the measurement of the invisible Z width at LEP I
\cite{sigma}.

As a result of the Yukawa coupling of the Higgs triplet to the leptons
$\ell \Delta \ell$ a Majorana neutrino mass matrix will be generated,
as follows 
\beq
\label{mnu} 
m_\nu = h_\nu v_3 
\eeq 
In order to comply with cosmological limits the magnitude of the
Yukawa couplings and/or $v_3$ is substantially restricted. We are
interested in the limit of small values of the triplet VEV $v_3$. In
this limit the smallness of neutrino mass in \eq{mnu} can be ascribed
to the smallness of $v_3$ without having to invoke tiny Yukawa
couplings. This is exactly what naturally happens in a seesaw scheme
\cite{LR}.

Taking into account the fact that this model contains one
doubly-charged and one singly-charged scalar boson, in addition to the
one charged unphysical \21 Goldstone mode (longitudinal $W^\pm$), it
follows that the neutral Higgs sector of this model is composed by
four real fields. Due to CP invariance they split into two unmixed
sectors. For example, for $v_3 \ll \kappa$ and $v_3 \ll v_2$ the
CP--even Higgs mass squared matrix may be written as
\begin{equation}
\label{MR}
{\bold M_R^2}=
\left[
\begin{array}{cc}
m_H^2 &  \left(\gamma m_H^2  - 2 m_A^2 \right) \frac{v_3}{ v_2}\\
\left(\gamma m_H^2  - 2 m_A^2 \right) \frac{v_3}{ v_2} & m_A^2  + \O(v_3^2)
\end{array}  \right]
\end{equation}
where $\gamma$ is a ratio of $\lambda$'s, $m_H^2$ is the SM Higgs
mass, and $m_A^2$ is the physical pseudoscalar boson mass given by
\begin{equation}
m_A^2=\half\kappa\frad{v_2^2+4v_3^2 }{v_3}
\label{MA}
\end{equation}
This mass results from diagonalizing the $2\times 2$ CP--odd Higgs
mass squared matrix via a rotation matrix $O_I$ defined by a small
angle 
\beq
\sin\beta = \frac{2v_3}{\sqrt(v_2^2 + 2 v_3^2)}
\eeq
and obeying $O_I{\bold M_I^2}O_I^T={\rm diag}(0,m_A^2)$, so that the
first state is the unphysical Goldstone boson.  On the other hand
$O_R{\bold M_R^2}O_R^T={\rm diag}(m_{H_1}^2,m_{H_2}^2)$ where we take,
by definition, $m_{H_1}\le m_{H_2}$, so that the first state
corresponds to the lightest CP--even Higgs boson.  The corresponding
matrix $O_R$ is given in terms of an angle $\alpha$. The parameter
$\sin\alpha$ determining the projection of $H_1$ along the triplet can
be large when $m_A < m_H$, as we will see below.

Note that none of the CP-even or CP-odd Higgs boson masses obtained in
our model lie at the scale $v_3$.  Consequently, even though the model
suffers from the usual hierarchy problem, this may be avoided by
supersymmetrization, as in the SM. This would estabilize all Higgs
masses at the weak scale.

The W and Z masses come from the kinetic part of the scalar Lagrangian
\begin{equation}
{\cal L}_0=({\cal D}_\mu\phi)^\dagger{\cal D}^\mu\phi
+{\rm tr}\,[({\cal D}_\mu\Delta)^\dagger{\cal D}^\mu\Delta]
\label{der}
\end{equation}
where the covariant derivative is defined by
\begin{equation}
{\cal D}=\partial^\mu+ig{\bf T}\cdot{\bf W}^\mu+\frad i2g'YV^\mu
\end{equation}
where $g$ and $g'$ are the $SU(2)$ and $U(1)$ gauge couplings
respectively.  The generators act on the scalars fields as 
\begin{eqnarray}
{\bf T}\phi&=\frad12{\vec\tau}\phi,\qquad
{\bf T}\Delta&=-\frad12{\vec\tau}\Delta-\frad12\Delta{\vec\tau}^*\nonumber\\
Y\phi&=-1\phi,\qquad Y\Delta&=2\Delta\,,
\label{W}
\end{eqnarray}
With these definitions we have $T_3\phi^0=\frad12\phi^0$ and
$T_3\Delta^0=-1\Delta^0$. The W mass is given by \cite{2227}
\begin{equation}
m_W^2=\frac{1}{4}g^2(v_2^2+2v_3^2)
\label{mw}
\end{equation}
so that $\sqrt{v_2^2 + 2v_3^2} \simeq 246$ GeV. From the measurement
of the $\rho$ parameter one has \cite{pichval97}
\begin{equation}
\rho=1+\frac{2v_3^2}{v_2^2+2v_3^2}=1.001\pm 0.002\,.
\label{rho}
\end{equation}
which implies in practice that $v_3\leq 9.5\:$GeV and $v_2$ almost
fixed, leading to $\sin\beta \lsim 10^{-2}$.  This restriction on
$v_3$ is automatically fulfilled for the values we are
dealing with.

\begin{figure}
\centerline{\protect\hbox{
\psfig{file=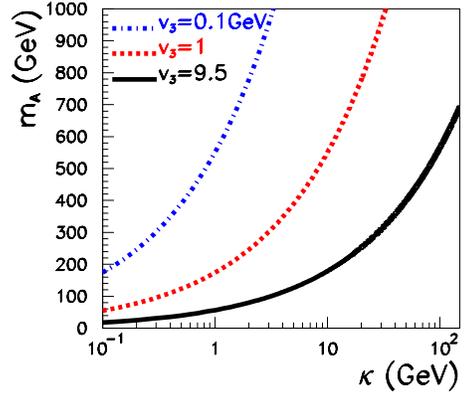,height=5.5truecm,width=7truecm}}}
\caption{Lightest CP--odd Higgs boson mass versus
$\kappa$ for different $v_3$ values. }
\label{ma-k}
\end{figure}

In order to have an idea of the expectations of the model for the
various Higgs boson masses we diagonalize the exact mass matrices and impose
the potential minimisation conditions, checking the positivity of the
physical CP-even and CP-odd eigenvalues.  In \fig{ma-k} we show the
CP--odd Higgs boson mass in our model versus $\kappa$ for different
$v_3$ values. The allowed region lies above the curve corresponding to
$v_3 \approx 9.5$ GeV.
\begin{figure}
\centerline{\protect\hbox{
\psfig{file=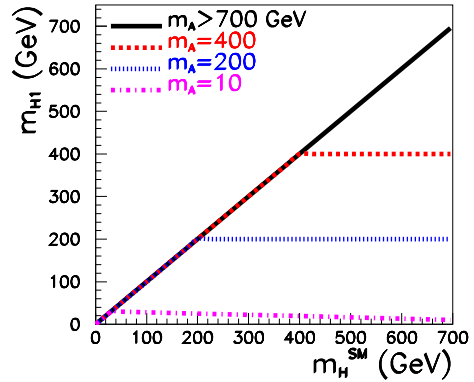,height=5.5truecm,width=7truecm}}}
\caption{Lightest CP--even Higgs mass vs the SM Higgs boson mass 
for different $m_A$ values.}
\label{mh-la1}
\end{figure}
Similarly in \fig{mh-la1} we show $m_{H_1}$, the mass of the lightest
CP--even Higgs boson in our model, as a function of the SM Higgs mass
$m_H$ for different $m_A$ values.  For example, if we fix $m_A$ at 200
GeV we find that $H_1$ lies along the solid diagonal line and is
almost the same as the SM doublet Higgs boson up to $m_H \sim
200$ GeV.  Past this value the $H_1$ mass lies on the horizontal line
at $m_{H_1} = 200$ GeV and $H_1$ is therefore mostly triplet. As we
will see later, it will decay mostly to neutrinos instead of to
b-quarks. Another feature worth noticing is that there can be a
substantial mixing in the CP--even sector and, as a result, the $H_1$
mass may also be lower than the SM Higgs mass $m_H$, especially for
small $m_A$ values.  Finally, we can also see that $m_{H_1} < m_A$, so
that the decay $H_1 \to A A$ does not occur.

The mechanisms for Higgs boson production at $e^+e^-$ colliders are
the emission of a CP--even Higgs by a $Z$--boson, and the associated
production of a CP--even Higgs and a CP--odd Higgs $A$. The relevant
couplings are given by
\begin{eqnarray}
{\cal L}_{H_aAZ}=\frad g{2c_w}Z^\mu\left[\,R_2
\stackrel{\longleftrightarrow}{\partial^\mu}\,I_2-2\,R_3
\stackrel{\longleftrightarrow}{\partial^\mu}\,I_3\right]\nonumber\\
=\frad g{2c_w}Z^\mu\left[
{\sin\beta}O^R_{a2}-2{\cos\beta}O^R_{a3}\right]\,H_a
\stackrel{\longleftrightarrow}{\partial^\mu}\,A.
\label{ZHAa}
\end{eqnarray}
where $c_W\equiv\cos\theta_W$ and $H_a$ is any of the two CP--even
neutral Higgs bosons. The parameter defined by
\begin{equation}
\epsilon_A={\sin\beta}\cos\alpha+2{\cos\beta}\sin\alpha \approx 2\sin\alpha
\end{equation}
will determine the strength of the $H_1AZ$ coupling. From \eq{der} we
find that the $H_a ZZ$ couplings are
\begin{equation}
{\cal L}_{H_a ZZ}={g m_Z \over{4c_W^2}} Z^\mu Z_\mu
\left[\cos\beta O^R_{a2}+2\sin\beta O^R_{a3}\right]H_a,
\end{equation}
and correspondingly we define 
\begin{equation}
\epsilon_B=\cos\beta\cos\alpha-2\sin\beta\sin\alpha \approx \cos\alpha
\label{eps_B}
\end{equation}
as a measure of the strength of the $H_1ZZ$ coupling. 

Notice the factor of 2 in the expressions for the parameters
$\epsilon_A$ and $\epsilon_B$ which determine the Bjorken and the
associated production cross sections, respectively. It comes from
the hypercharge of the triplet.

We now turn to the couplings relevant for the invisible decay of the
lightest Higgs bosons. The pseudoscalar $A$ may decay into stable
neutrinos, via the triplet Yukawa coupling $h_\nu$ of \eq{mnu}, or
into $ b\bar{b}$, via its projection onto the doublet. In order to
evaluate the relative importance of the two branchs we need
information on the neutrino mass. In the presence of the cubic
lepton--number--breaking term $\kappa$ there is no efficient means of
reducing the relic neutrino number density, as a result of which
neutrinos in this model must obey the limit from cosmology on stable
neutrino masses
\begin{equation}
m_\nu\lsim 92\: \Omega h^2 \hbox{eV}
\label{cosmo}
\end{equation} 
where $\Omega h^2 \leq 1$ is a basic cosmological parameter \cite{omega}.
Present determinations give $h$ is $.65 \pm 0.1$, while the total $\Omega$
may still be as large as one, as suggested by inflationary models.
 
Without adding singlet scalar bosons there is, on the other hand,
no way to introduce the majoron in a phenomenologically acceptable 
way, which does not conflict with LEP I data on the observed width of
the $Z$ into invisible channels. From \eq{mnu} we see that
\beq
\label{0-}
\frac{\Gamma(A\to b{\bar b})}{\Gamma(A\to \nu{\bar \nu})}
=\left(\frac{3 h_b\sin\beta}{h_\nu\cos\beta}\right)^2
\approx \left(\frac{6m_b}{m_\nu} \right)^2 \frac{v_3^4}{ v_2^4}
\eeq
where $h_\nu$ here denotes the Yukawa coupling of the most massive of
the neutrinos with the triplet. One sees that for $m_\nu \sim 10$ eV
and $v_3 \lsim$ few MeV, the decay of $A$ to neutrinos will be
dominant.

An analogous calculation for the CP--even sector gives
\beq
\label{0+}
\frac{\Gamma(H_1 \to b{\bar b})}{\Gamma(H_1 \to \nu{\bar \nu})}
= \left(\frac{3 h_b\cos\alpha}{h_\nu\sin\alpha}\right)^2 %%\nonumber\\
\lsim \left(\frac{6m_b}{m_\nu} \right)^2 \frac{v_3^4}{ v_2^4} \gamma^2
\eeq 
Clearly for $\cos \alpha \to 0$ $H_1$ is mostly triplet and therefore
decays mainly to neutrinos. This corresponds to the horizontal lines
in Fig. 2. In the opposite situation $\cos \alpha \to 1$ $H_1$ is
mostly doublet and decays mainly to $b\bar{b}$. The price we must pay
in order to have $\cos \alpha \approx 0$ is to have again a small
$v_3$ as seen from \eq{0+}.  In the last step in \eq{0+} we assumed
$m_A \ll m_H$ in order to obtain a conservative upper bound on $v_3$
as a function of $\gamma$.  Again, one sees that for $m_\nu \sim 10$
eV and $\gamma=1$ we find that for $v_3$ values in the MeV range the
lightest CP--even Higgs boson $H_1$ will also decay invisibly. For
smaller $\gamma$ values the upper bound on $v_3$ is relaxed
correspondingly.

Note that in the simplest scheme presented above the smallness of
$v_3$ is put in by hand. However, the simplest model may be regarded
as an effective parametrization of a more complete left-right
symmetric see-saw scheme \cite{LR} in which lepton number is a local
symmetry violated spontaneously at a large scale $v_R$.  The smallness
of $v_3$ would account for the smallness of neutrino mass and would
arise naturally from a minimization condition of the scalar boson
potential leading to $v_3 \sim m_W^2/v_R$.

We now perform a model independent study of the limits that can be set
based on Higgs boson production in $e^+e^-$ colliders at the $Z$ peak
and its subsequent invisible decay.  Consider the massive pseudoscalar
$A$ and the lightest CP--even scalar $H_1$. As we have seen $H_1$ and
$A$ may decay invisibly when $v_3$ is small. As seen above, for small
$\gamma$ the bound on $v_3$ for $H_1$ to decay invisibly may be
somewhat relaxed.  The basic parameters needed to describe the
implications of the production of Higgs bosons at the $Z$ peak in this
model are the masses $m_A$ and $m_{H_1}$, the coupling parameters
$\epsilon_A$ and $\epsilon_B$ which determine the Bjorken and
associated production cross sections and the product of the visible
and invisible decay branching ratios.

The Bjorken process contribution to the invisible $Z$ width $Z
\to Z^* H_1$ is, for most of the parameter space, very small
compared to that of the associated production.  Thus, in order to get
a conservative bound, we only consider the associated process, bearing
in mind that the inclusion of the Bjorken contribution would only
improve our results, i.e., would exclude a slightly wider region of
parameter space.

Therefore we consider in what follows the limits that can be set on
associated Higgs boson production at the Z peak, $e^+e^- \to Z \to H_1
A$ when both CP-even ($H_1$) as well as CP-odd Higgs bosons ($A$)
decay invisibly.  One can write the contribution to the
invisible $Z$ width as:
%
%%\begin{equation}
$$
\Delta \Gamma_{inv} = \frac{{\epsilon_A}^2}2 
\lambda^{\frac32}\left(1,{m_{H_1}^2 \over m_Z^2},{m_A^2 \over m_Z^2}\right) 
\Gamma(Z \to \nu_e \bar{\nu_e})B_{inv}A_{inv} \nonumber
%%\end{equation}
$$
where $B_{inv}$ and $A_{inv}$ denote the invisible branching ratios of
$BR(H_1 \to \nu{\bar \nu})$ and $BR(A \to \nu{\bar \nu})$, respectively,
and $\lambda$ is the usual K\"allen function 
$\lambda(a,b,c)=(a-b-c)^2-4bc$.

\begin{figure}
\centerline{\protect\hbox{
\psfig{file=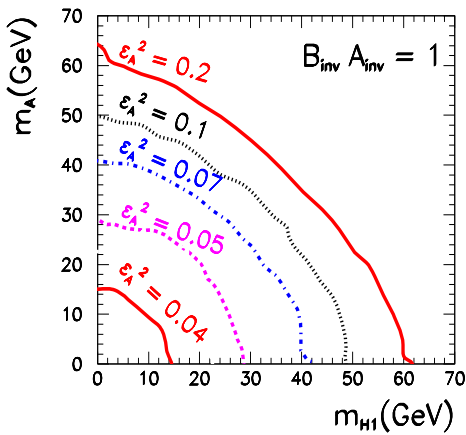,height=6truecm,width=7truecm}}}
\caption{95 \% CL bounds on ${\epsilon_A}^2$ in $m_H$-$m_A$ plane 
when both $H_1$ and $A$ decay to neutrinos.}
\label{results1}
\end{figure}

Taking into account the experimental error in the determination of the
invisible $Z$ width, given by $\sigma=2$ MeV \cite{sigma}, we have
determined 95 \% CL bounds on ${\epsilon_A}^2$ in the $m_{H_1}$-$m_A$
plane for fixed values of the product $B_{inv}A_{inv}$.  As it is well
known, the integration of a Gaussian probability distribution from
$-\infty$ to $+1.64 \sigma$ gives 0.95, so that the 95 \% CL exclusion
region is defined by imposing $\Delta \Gamma_{inv} > 1.64 \sigma$. In
\fig{results1} we show these results for a fixed value of the product
$B_{inv}A_{inv} =1$. This corresponds to both scalar and pseudoscalar
decaying totally to neutrinos. This choice is meant for
definiteness. The constraints corresponding to any other value of
$B_{inv}A_{inv}$ may be simply obtained by rescaling the results for
our reference value given in \fig{results1}. In the plot we have five
curves labeled by a value of $\epsilon_A^2$.
No points below each of these curves are allowed with $\epsilon_A^2$
larger than that value. We see from this plot that simply by using the
neutrino counting at the Z peak one can already impose important
constraints on the parameters of the model.  For example, for $H_1$
and $A$ masses around 20 GeV the upper bound on ${\epsilon_A}^2$ is a
few times $10^{-2}$. Going beyond this requires a dedicated analysis
of the various event topologies that are possible in this model, for
example, di-jet plus missing momentum. Fortunately the same topologies
are also the ones present in other models where the invisible Higgs
boson decay involves majorons, considered both for LEP I as well as
LEP II data \cite{ebolep}. The results of that analysis can easily be
adapted to the present model. In any case an updated analysis of the
present LEP II data by the LEP colaborations themselves would be
welcome.

In short, we have illustrated, with a very simple model, how LEP
experiments may shed information on the Higgs boson sectors of models
of neutrino mass.  In contrast, all previously considered models with
invisibly--decaying Higgs bosons invoked the existence of majorons and
employed only physics at the weak scale.  We have shown how the
invisibly decaying Higgs boson signal can arise from the existence of
a small scale ($v_3$ in our model) associated to neutrino masses in a
model with explicit violation of lepton number.  In a more complete
see-saw left-right scheme, the smallness of $v_3$ would arise
naturally as  $v_3 \sim m_W^2/v_R$. 

This work was supported by DGICYT under grant PB95-1077, and by the
TMR network grant ERBFMRXCT960090 of the European Union. M.A.D.  was
supported by a DGICYT postdoctoral grant, while M.A.G. and D.R. were
supported by pre-doctoral grants from DGICYT and COLCIENCIAS respectively.
We thank Rocky Kolb and Sasha Dolgov for discussions on the present
value of $\Omega h^2$ \cite{omega}.

\end{document}